# Effect of Different Device Parameters on Tin Based Perovskite Solar Cell Coupled with In$_2$S$_3$ Electron Transport Layer and CuSCN and Spiro-OMeTAD Alternative Hole Transport Layers for High Efficiency Performance

*Intekhab Alam[1*], Md Ali Ashraf[2]*

[1]Department of Mechanical Engineering (ME), Bangladesh University of Engineering and Technology (BUET), East Campus, Dhaka-1000, Bangladesh

[2]Department of Industrial and Production Engineering (IPE), Bangladesh University of Engineering and Technology (BUET), East Campus, Dhaka-1000, Bangladesh

*Corresponding author
Email: intekhabsanglap@gmail.com

# Effect of Different Device Parameters on Tin Based Perovskite Solar Cell Coupled with $In_2S_3$ Electron Transport Layer and CuSCN and Spiro-OMeTAD Alternative Hole Transport Layers for High Efficiency Performance

**Abstract**

SCAPS 1-D was used for the simulation of lead-free environmentally benign methylammonium tin-iodide ($CH_3NH_3SnI_3$) based solar cell. Indium sulphide ($In_2S_3$) was utilized as the electron transport layer (ETL) for its high carrier mobility and optimized band structure, unlike traditional titanium oxide ($TiO_2$) ETL. Traditional expensive spiro-OMeTAD ($C_{81}H_{68}N_4O_8$) and cheaper cuprous thiocyanate (CuSCN) were utilized alternatively as hole transport layer (HTL) to observe the effect of different HTL on cell performance. We investigated the trend in electrical measurements by altering parameters such as thickness, defect density, valence band (VB) effective density of state and bandgap of the absorber layer, interfacial trap densities and defect density of ETL. At optimum condition, the device revealed the highest efficiency of 18.45% for CuSCN (HTL) and 19.32% for spiro-OMeTAD (HTL) configuration. The effect of working temperature, the wavelength of light and band-to-band radiative recombination rate was also observed for both configurations. All these simulation results will help to fabricate eco-friendly high-efficiency perovskite solar cell by replacing the commonly used toxic lead-based perovskite.

## 1. Introduction

Clean, green and sustainable energy sources like solar cells are crucial to compensate for the ever-increasing energy demand because relying on non-renewable sources of energy is not a feasible solution. Energy generation from the solar cells has been mostly considered among other renewable energy sources due to its low costs of distributed power generation, operation and maintenance (Kharaji Manouchehrabadi and Yaghoubi 2020). The net solar generation is increasing at a rate of about 8.3% annually (Ashraf and Alam 2020). Highly expensive silicon-based solar cells have an efficiency of 12%-17.5% only, but they have been dominating the solar market for many years with 94% market share (Husainat, Ali et al. 2020). So, an extensive search for alternative solar materials has become a necessity. Thin-film technology has made drastic changes in the solar cell industry in the 2000s onwards by improving the efficiency of the solar cell up to 21% and making the solar cells lighter, thin and durable (Bangari, Singh et al. 2020). Dye-sensitized solar cells have induced intensive interests over the past decades due to its low cost and simple preparation processes (Lan, Wu et al. 2014). The uses of the new generation materials and devices such as dye-sensitized, thin-film cadmium telluride and thin-film copper indium gallium selenide based solar cells have been on the rise to offset the silicon-based market, but some of them contain expensive and toxic elements (MaríSoucase, Pradas et al. 2016).

Over the years, perovskite solar cells have gained the attention of the researchers for having low processing cost, solution processing and excellent light-harvesting characteristics along with an efficiency reaching up to 25.2% (Mandadapu, Vedanayakam et al. 2017, Tai, Cao et al. 2019). Hybrid organic-inorganic metal halide perovskite solar cells have a tunable bandgap, lower excitation binding energy, excellent optoelectrical properties, long carrier diffusion distance and outstanding performance almost equivalent to silicon-based solar cells (Zhou, Zhou et al. 2018, Husainat, Ali et al. 2020, Meng, Chen et al. 2020). Among them, methylammonium lead-iodide ($MAPbI_3$) has achieved over 21% efficiency and is used most commonly as absorber layer due to its excellent efficiency, regardless of its stability problem (Chen, Turedi et al. 2019). However, due to the toxic nature of lead, methylammonium tin-iodide ($MASnI_3$) is a suitable replacement. Moreover, $MASnI_3$ has a lower bandgap around 1.25 to 1.3 eV, an absorption coefficient around $10^5$ and excellent optical properties along with high mobility and small effective mass (Lee, He et al. 2012, Iefanova, Adhikari et al. 2016, Anwar, Mahbub et al. 2017). It has achieved 7.13% efficiency, which is significantly low compared to its lead-based counterpart (Li, Zhang et al. 2019). This low-efficiency is because $MASnI_3$ is susceptible to the atmosphere due to facile oxidation and forms environmentally benign $Sn^{4+}$ from $Sn^{2+}$. However, this transformation reduces its photovoltaic performance by causing significant carrier recombination (Mandadapu, Vedanayakam et al. 2017, Tai, Cao et al. 2019). Its self-doping problem due to oxidation can be solved most effectively by the inclusion of tin-halide ($SnF_2$ or $SnCl_2$) resulting in improved stability and performance (Tai, Cao et al. 2019). Two main deposition techniques applied in the construction of high-grade perovskite thin films are vapour-based and solution-based deposition (Husainat, Ali et al. 2020).

The high efficiency achieved by using organic 2,2′,7,7′-Tetrakis[N, N-di(4-methoxyphenyl)amino]-9,9′-spirobifluorene or spiro-OMeTAD ($C_{81}H_{68}N_4O_8$) as the hole transport layer (HTL) results from the combination of reasonable charge carrier mobility added with its amorphous nature and high solubility, enhanced glass-forming capacity and high electrochemical stability (Bach, Lupo et al. 1998, Fantacci, De Angelis et al. 2011). Because of its high efficiency, it is mostly used as HTL regardless of its high cost and thermal instability (Arora, Dar et al. 2017). However, inorganic cuprous thiocyanate (CuSCN) can be a cheaper replacement of expensive and degradable organic spiro-OMeTAD (Yang, Wang et al. 2019). CuSCN exhibits excellent hole mobility, well-aligned work function and material stability, but more inferior photovoltaic property than organic HTL as it causes damage to the perovskite layer during coating (Arora, Dar et al. 2017, Lazemi, Asgharizadeh et al. 2018, Yang, Wang et al. 2019). It can be fabricated by the doctor blade method, spin coating and spray deposition (Yang, Wang et al. 2019).

Electron transport layer (ETL) should have proper band alignment to facilitate electron transport, excellent carrier mobility and wide bandgap (Islam, Jani et al. 2020). Usually, titanium oxide ($TiO_2$) is used as ETL in the perovskite solar cells which has limitations due to its high-temperature fabrication, intrinsic slow electron mobility and can cause a disturbance in charge transport (Lee, He et al. 2012, Iefanova, Adhikari et al. 2016, Anwar, Mahbub et al. 2017). Indium sulphide ($In_2S_3$) is a suitable replacement of traditional $TiO_2$ as ETL due to its higher carrier mobility, good stability, optimized band structure and enhanced light tapping and also, it can even outperform $TiO_2$ when used in perovskite solar cells (Hou, Chen et al. 2017, Xu, Wu et al. 2018, Yu, Zhao et al. 2019). Photochemical deposition, spray pyrolysis, thermal evaporation and modulated flux deposition are conventional techniques for the production of $In_2S_3$ (ETL) (Hossain 2012). Furthermore, fluorine-doped tin oxide (FTO) has tunable bandgap, high transparency in UV/ IR spectrum, high electrical conductivity, high chemical stability, proper surface texture for increasing light scattering and absorption (Huang, Ren et al. 2014), and can be conveniently used as the window layer. FTO is most efficiently fabricated by spray pyrolysis (Aouaj, Diaz et al. 2009), and can enhance the stability of $MASnI_3$ (Mandadapu, Vedanayakam et al. 2017).

In this simulation, the aim was to predict the best possible efficiency of soda-lime glass (*SLG*)/ *FTO*/ *$In_2S_3$*/ *$MASnI_3$*/ *(Spiro-OMeTAD, CuSCN)*/ *gold (Au)* solar cell by optimally varying the thickness, defect density, bandgap, valence band (VB) effective density of state of the absorber layer, defect density of ETL and trap densities of interfacial layers and also, to investigate the effects of change of temperature, the wavelength of light and band-to-band radiative recombination rate. Although there are some experimental works regarding $MAPbI_3$-based solar cell with $In_2S_3$ as ETL and spiro-OMeTAD as HTL (Xu, Wu et al. 2018, Yu, Zhao et al. 2019), there are toxicity issues regarding them. So, we are proposing a novel simulated model replacing the toxic lead-based perovskite with tin-based $MASnI_3$ to observe its performance, and also, suggesting CuSCN as an alternative HTL because of its low-cost fabrication and excellent mobility.

## 2. Methodology

All the simulations were performed by SCAPS-1D software, which was developed at the Department of Electronics and Information Systems of Gent University, Belgium (Niemegeers, Burgelman et al. 2014). SCAPS uses analytical physics of solar cell device including transport mechanism, individual carrier current densities, electric-field distribution and recombination profile. The software numerically solves the Poisson's and continuity equations that are needed for charge carrier transport, which are given in the Supplementary Document. The physical quantities including open-circuit voltage ($V_{OC}$), short-circuit current density ($J_{SC}$), fill factor (FF) and photoconversion efficiency (PCE) can be calculated in light and dark medium and also at different illuminations and temperatures. The good agreements between the experimental results and SCAPS simulation results motivated us to use the simulation tool in this work (Burgelman, Nollet et al. 2000, Khelifi, Verschraegen et al. 2008, Marlein, Decock et al. 2009).

### 2.1 Architecture of the Devices

The structure of soda-lime glass (SLG)/ FTO/ $In_2S_3$/ $MASnI_3$/ (Spiro-OMeTAD, CuSCN)/ gold (Au) solar cell is shown at Figure 1(a) which had two configurations (Anwar, Mahbub et al. 2017, Yu, Zhao et al. 2019), where spiro-OMeTAD and CuSCN were used alternatively as two different p-type HTLs over the Au back contact, $MASnI_3$ was a p-type absorber layer, $In_2S_3$ was an n-type ETL, and FTO was an n-type window layer. It was a solid-state planar heterojunction p-i-n solar cell with p-type $MASnI_3$ sandwiched between the n-type ETL and p-type HTL. SLG was over FTO and light fell on the SLG side. Both arrangements had two interfacial layers- HTL/ absorber layer and absorber layer/ ETL. Bandgap alignment of the device is shown in figure 1(b).

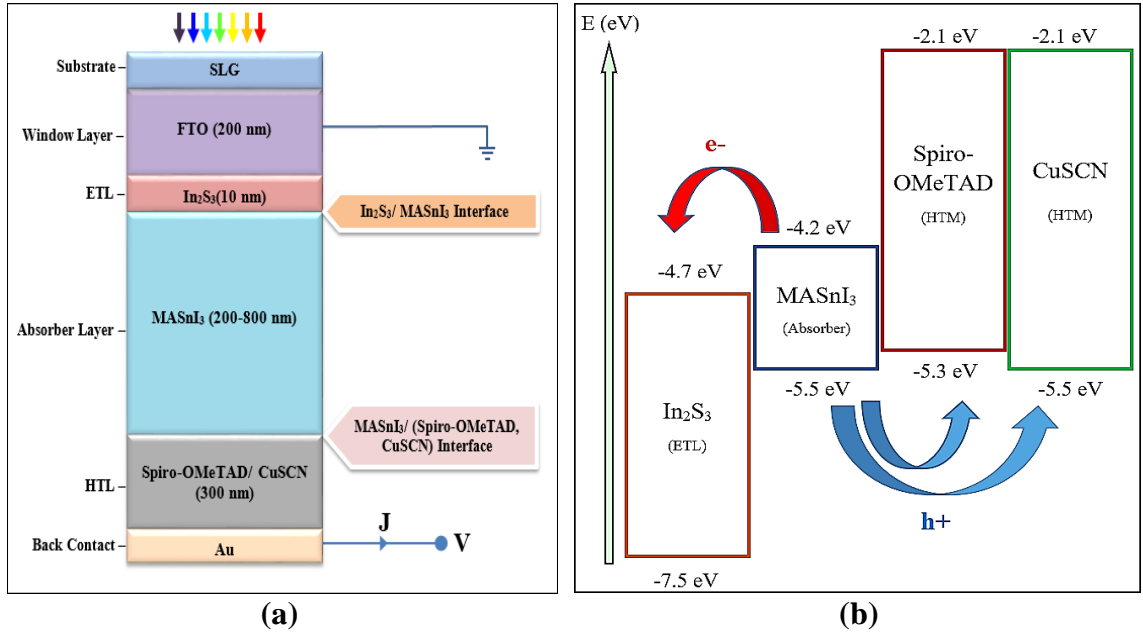

(a) (b)

**Figure 1**: **(a)** Device architecture; **(b)** Bandgap alignment for tin-based perovskite solar cell with both HTLs.

*2.2 Simulated parameters*

The material parameters used in the simulation are listed in table 1. The values were meticulously chosen from literature, and the previous simulation works (Lenka, Soibam et al. , Hossain, Chelvanathan et al. 2011, Anwar, Mahbub et al. 2017, Hou, Chen et al. 2017, Xu, Wu et al. 2018, Ashraf and Alam 2020). The default value of temperature was taken as 300K with standard illumination of AM1.5G and was varied later. All the simulations were run at light-medium, 1000 Wm$^{-2}$ illumination, 1 Ohm.cm$^2$ series resistance and 1000 Ohm.cm$^2$ shunt resistance. Electron and hole thermal velocities were kept constant at $10^7$ cms$^{-1}$. Band-to-band radiative recombination coefficient value was taken as $3\times10^{-11}$ cm$^3$sec$^{-1}$ initially, which was varied later and also, both of auger electron and hole capture coefficients were fixed at $1\times10^{-29}$ cm$^6$sec$^{-1}$ (MaríSoucase, Pradas et al. 2016). Neutral type and donor type defects were considered for HTLs and absorber layer, respectively. For both ETL and window layer, acceptor type defects were considered. Absorber layer and ETL had variable total defect density ($N_t$), whereas HTLs had fixed $N_t$. All interfacial layers (HTL/ absorber layer and absorber layer/ ETL) had neutral type defects and variable trap densities. For Au back contact, work function was 5.1 eV, and thermionic emission/ surface recombination velocities for electrons and holes were $10^5$ cms$^{-1}$ and $10^7$ cms$^{-1}$, respectively. The optical absorption constant, α(hν), for the perovskite layer was set by the new "Eg-sqrt" model, and the details of the model were shown in the Supplementary Document (Ashraf and Alam 2020, Niemegeers 2014, Du, Wang et al. 2016, Sharma 2019).

**Table 1**: SCAPS-1D input material parameters used in the solar cell simulation.

| | Units | Window Layer | ETL | Absorber Layer | HTL | |
|---|---|---|---|---|---|---|
| **Material** | | FTO | $In_2S_3$ | $MASnI_3$ | Spiro-OMeTAD | CuSCN |
| **Thickness** | nm | 200 | 10 | 200-800 | 300 | 300 |
| **Bandgap ($E_g$)** | eV | 3.20 | 2.80 | 1.30 | 3.20 | 3.40 |
| **Electron affinity ($\chi$)** | eV | 4.40 | 4.70 | 4.20 | 2.10 | 2.10 |
| **Relative Permittivity ($\varepsilon_r$)** | - | 9.00 | 13.50 | 10.00 | 3.00 | 10.00 |
| **CB effective density of states ($N_c$)** | $cm^{-3}$ | 2.2E+18 | 1.8E+19 | 1.0E+18 | 2.5E+18 | 2.5E+18 |
| **VB effective density of states ($N_v$)** | $cm^{-3}$ | 1.8E+19 | 4.0E+13 | 1.0E+18 | 1.8E+19 | 1.8E+19 |
| **Electron mobility ($\mu_n$)** | $cm^2\,V^{-1}\,s^{-1}$ | 20 | 400 | 1.6 | 2.0E-4 | 2.0E-4 |
| **Hole mobility ($\mu_p$)** | $cm^2\,V^{-1}\,s^{-1}$ | 10 | 210 | 1.6 | 2.0E-4 | 100E-2 |
| **Donor density ($N_d$)** | $cm^{-3}$ | 1.0E+18 | 1.0E+17 | 0 | 0 | 0 |
| **Acceptor density ($N_a$)** | $cm^{-3}$ | 0 | 1.0E+1 | 3.2E+15 | 1.0E+20 | 1.0E+18 |
| **Defect density ($N_t$)** | $cm^{-3}$ | 1.0E+15 | 1.0E+18 | 4.5E+16 | 1.0E+14 | 1.0E+14 |

## 3. Result and Discussion

### 3.1 Effect of Absorber Layer and ETL Thickness

Estimation of absorber layer thickness is fundamental to determine the efficiency of a solar cell. At the start of the simulation, the thickness of the absorber layer ($MASnI_3$) was varied from 200 nm to 800 nm for both cell configurations to find out the influence of absorber layer thickness in cell performance while keeping other material parameters constant. The trends in simulated parameters such as PCE, FF, $V_{OC}$ and $J_{SC}$ with varying absorber layer thickness are shown in figure 2.

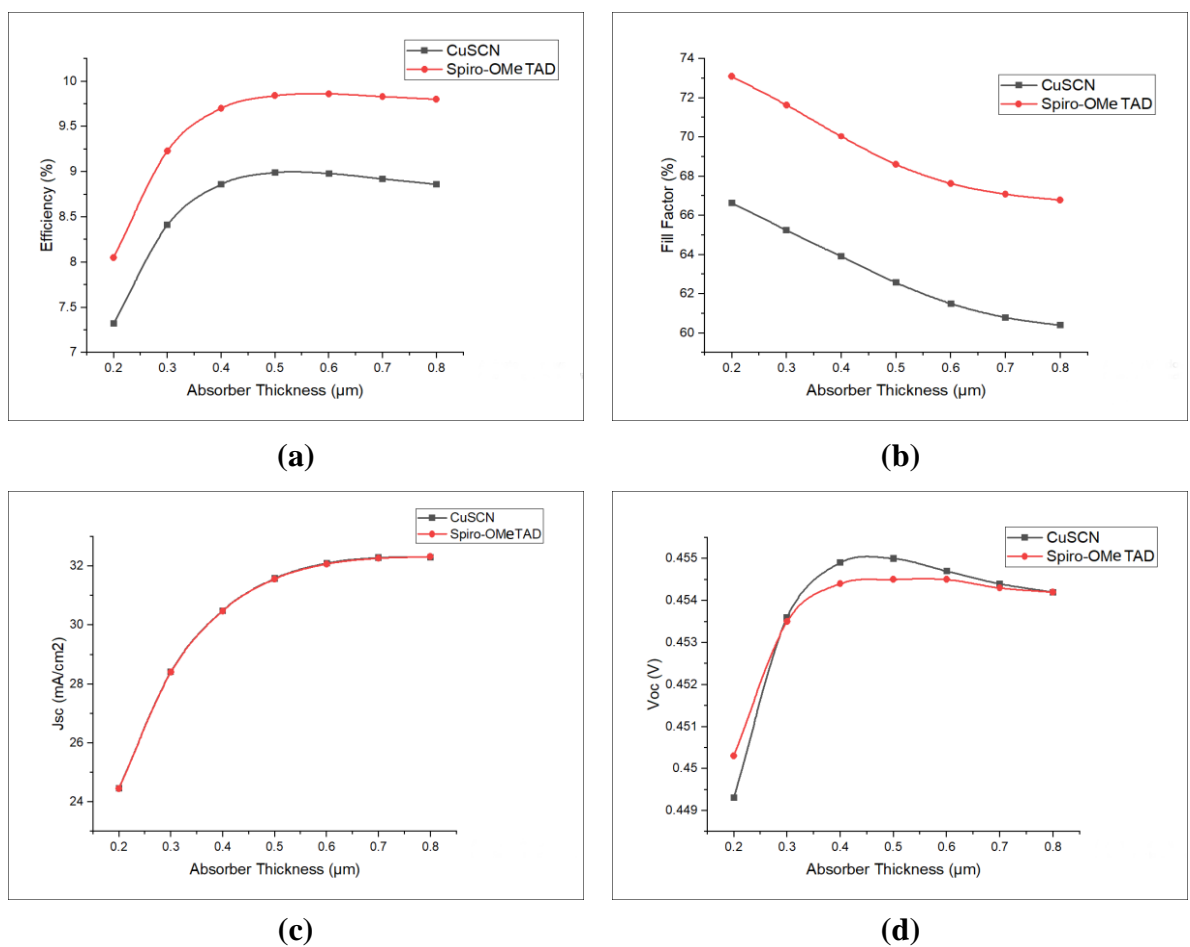

**Figure 2**: **(a)** PCE, **(b)** FF, **(c)** $J_{SC}$ and **(d)** $V_{OC}$ diagram with varying absorber layer thickness.

For CuSCN (HTL) arrangement, maximum PCE of 8.99% ($J_{SC}$ = 31.5794 mA/cm$^2$, FF = 62.57%, $V_{OC}$ = 0.455V) was found at 500 nm of absorber layer thickness and for spiro-OMeTAD (HTL) configuration, maximum PCE of 9.86% ($J_{SC}$ = 32.0715 mA/cm$^2$, FF = 67.63%, $V_{OC}$ = 0.4545 V) was found at 600 nm of absorber layer thickness. The thicknesses were in good agreement with the literature (Mandadapu, Vedanayakam et al. 2017, Baig, Khattak et al. 2018). However, both of the PCEs found for the best absorber thickness were significantly low compared to the 15.5% PCE of the FTO/ c-TiO$_2$/ m-TiO$_2$/ MAPbI$_3$/ Spiro-OMeTAD/ Au solar cell (De Los Santos, Cortina-Marrero et al. 2020). In both cases, efficiency and $V_{OC}$ increased up to a certain thickness, reaching the maximum value, then got reduced. Generally, $J_{SC}$ followed an upward trend, and FF followed a downward trend. $J_{SC}$ increases with the increasing thickness because a thicker absorber layer will absorb more photons resulting in creating more electron-hole pair. However, when the absorber thickness was lower than 300 nm, $J_{SC}$ declined at a faster rate which was due to enhanced recombination near the Au back contact. $J_{SC}$, reverse saturation current density ($J_0$) and $V_{OC}$ have a relation which is shown by the following equation: (Husainat, Ali et al. 2020)

$$V_{OC} = \frac{AK_BT}{q}\left[ln\left(1+\frac{J_{SC}}{J_0}\right)\right] \qquad (1)$$

Where, $V_{OC}$ (V) is the open-circuit voltage, $A$ is the ideality factor, q is an elementary charge, $\frac{K_BT}{q}$ (V) is the thermal voltage, $J_{SC}$ (mA/cm²) is the solar cell light generated current density and $J_0$ (mA/cm²) is the reverse saturation current density. According to equation (1), $J_0$ and $V_{OC}$ have an inverse relationship. Thinner absorber layer causes less electron-hole recombination that keeps the value of $J_0$ low, and that is why $V_{OC}$ had higher value initially. Then, with the increase of absorber thickness, the value of $J_0$ rises, causing the value of $V_{OC}$ to decline (Anwar, Mahbub et al. 2017). FF decreased with the increasing absorber thickness due to the increase in charge pathway resistance. Finally, PCE depends on $J_{SC}$, FF and $V_{OC}$ (Islam, Jani et al. 2020). Increasing absorber layer thickness enhances the generation of electron-hole pairs causing the rise in PCE initially. However, the chance of radiative recombination and charge pathway resistance enhances at the same time, causing the declination in PCE for thicker absorber layer. The external quantum efficiency (EQE) curves for the best absorber thickness for both configurations within the wavelength between 400-1000 nm are given in figure 3. The curves follow similar trends as the literature (Niemegeers 2014). It was found that near 400 nm wavelength, the EQE was slightly lower than 100% for both configurations and then, decreased with the increasing wavelength.

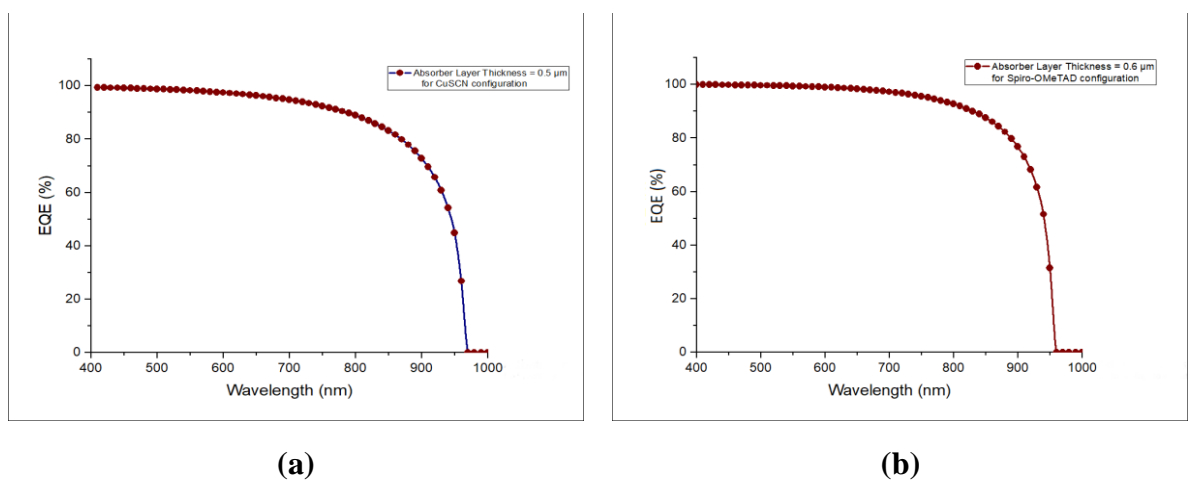

(a)        (b)

**Figure 3**: EQE curves between 400-1000 nm wavelength for the best absorber layer thickness for (a) CuSCN and (b) spiro-OMeTAD configuration.

## 3.2 Effect of Defect Density ($N_t$) of the Absorber Layer and ETL

In the simulation, defect density ($N_t$) of absorber layer was varied between $10^{10}$ cm$^{-3}$ to $10^{18}$ cm$^{-3}$ for the best absorber thickness to find the variation in efficiency, FF, $V_{OC}$ and $J_{SC}$ as shown in figure 4 for the solar cell with both configurations.

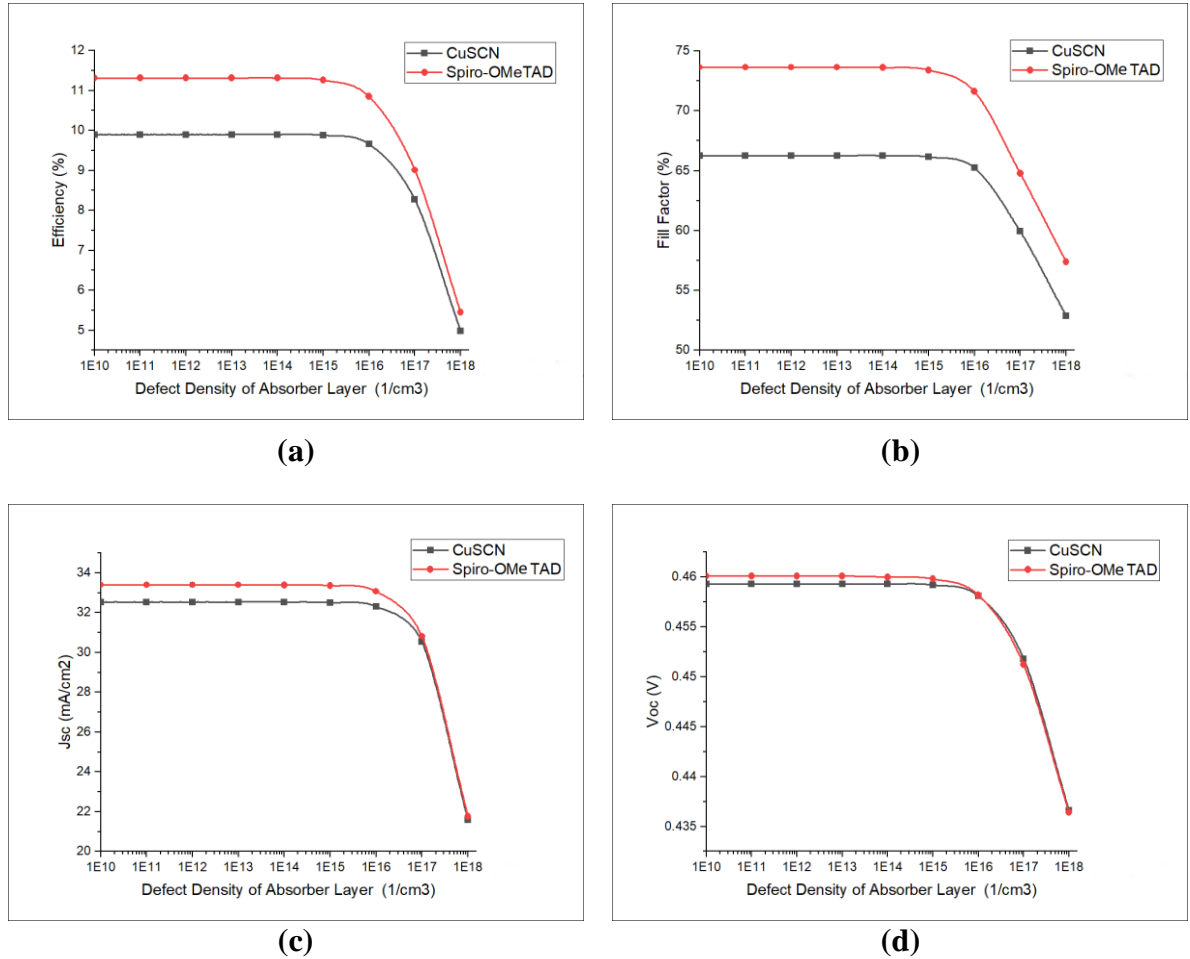

**Figure 4**: **(a)** PCE, **(b)** FF, **(c)** $J_{SC}$ and **(d)** $V_{OC}$ diagram with varying absorber defect density for the best absorber layer thickness.

With the increasing $N_t$ of the absorber layer, efficiency, FF, $V_{OC}$ and $J_{SC}$ of the solar cell decreased for both cases of HTLs. Nevertheless, with the decreasing $N_t$, efficiency got stable at a certain point for each case which was taken as the best defect density of the absorber layer. The best absorber defect density for both cases was $10^{14}$ cm$^{-3}$ same as the literature (Mandadapu, Vedanayakam et al. 2017, Nine, Hossain et al. 2019). For CuSCN as HTL, maximum PCE of 9.9% ($J_{SC}$ = 32.5366 mA/cm$^2$, FF = 66.26%, $V_{OC}$ = 0.4593 V) and for spiro-OMeTAD as HTL, maximum PCE of 11.31% ($J_{SC}$ = 33.3917 mA/cm$^2$, FF = 73.62%, $V_{OC}$ = 0.46 V) were found at $10^{14}$ cm$^{-3}$ absorber defect density. The external quantum efficiency (EQE) curves for the best absorber defect density at the best absorber thickness for both configurations are given in figure 5.

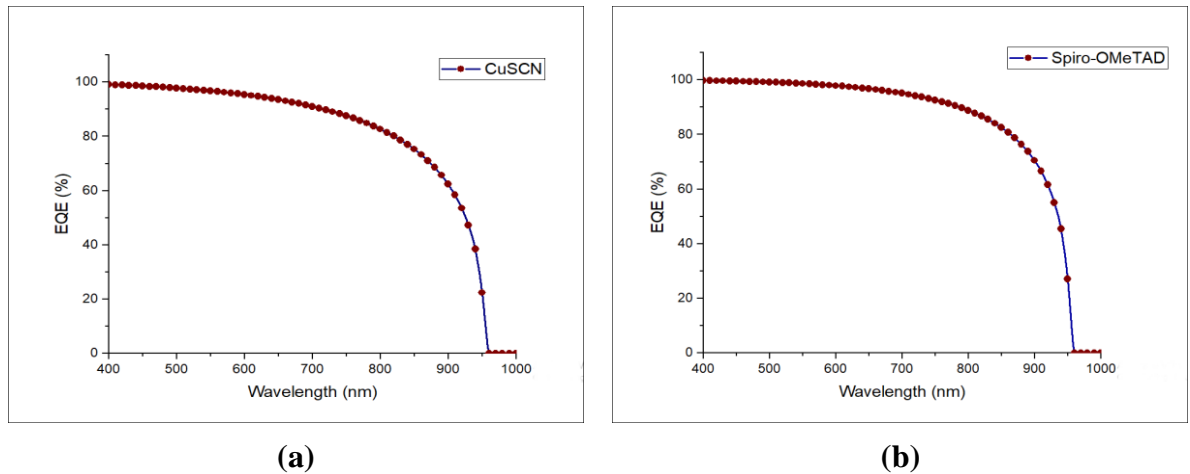

**(a)** **(b)**

**Figure 5**: EQE curves between 400-1000 nm wavelength for the best absorber defect density at the best absorber layer thickness for (a) CuSCN and (b) spiro-OMeTAD configuration.

Additional carrier recombination centres can be introduced by the defect states in the absorber layer increasing the recombination process of photo-generated carriers and resulting in the rise of $J_0$ and also, the reduction of diffusion length, $V_{OC}$ and $J_{SC}$, which in turn reduces PCE (Lin, Lin et al. 2014). The recombination centres at the deep energy levels are known as Shockley-Read-Hall non-radiative (SRH) recombination centre which clearly explains the effect of defect density of a perovskite absorber layer and Gaussian distributions can be incorporated in the absorber layer to quantitatively analyze the influence of defect states on solar cell performance (Wetzelaer, Scheepers et al. 2015, MaríSoucase, Pradas et al. 2016, Mandadapu, Vedanayakam et al. 2017). Corresponding equations of SRH recombination and Gaussian distributions are given in the Supplementary Document.

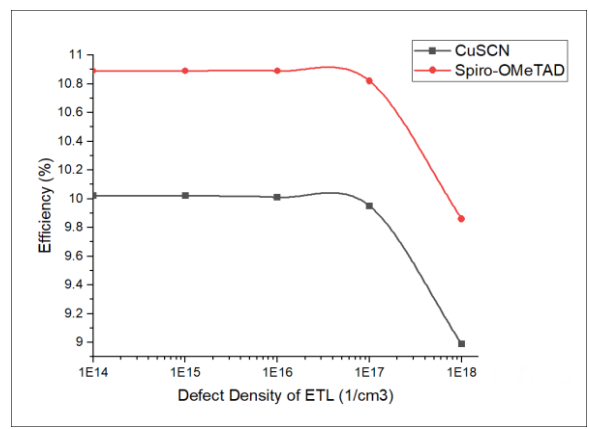

**Figure 6**: PCE diagram with varying ETL defect density for the best absorber layer thickness.

Similarly, the defect density of ETL was varied between $10^{14}$ cm$^{-3}$ to $10^{18}$ cm$^{-3}$ for the best absorber thickness to find the variation in PCE, as shown in figure 6 for both configurations. With the increasing $N_t$ of the ETL, PCE of the solar cell reduced for both cases. Moreover, when the value of defect density was being lowered, efficiency got stable at $10^{15}$ cm$^{-3}$, which was taken as the optimum defect density of the In$_2$S$_3$ ETL for both arrangements and was comparable to the optimum defect density found for TiO$_2$ ETL in the literature (Jeyakumar, Bag et al. 2020, Rai, Pandey et al. 2020). Cell with CuSCN (HTL) showed a maximum PCE of 10.02%, and for spiro-OMeTAD (HTL), maximum PCE of 10.89% was found at the optimum value of the defect density of ETL. The change of PCE with the change of the defect density of ETL was far less drastic than the defect density of absorber layer (almost 5.69 times less for spiro-OMeTAD and approximately 4.78 times less for CuSCN configuration). Similar to the absorber defect density, additional carrier recombination centres can be introduced by the defect states in the ETL, which causes a reduction in PCE.

### 3.3 Effect of Valence Band (VB) Effective Density of state ($N_V$) of Absorber Layer

Valence band (VB) effective density of state ($N_V$) of the absorber layer was varied from $10^{17}$ cm$^{-3}$ to $10^{19}$ cm$^{-3}$ for the best absorber thickness to find the variation in PCE as shown in figure 7 for both configurations.

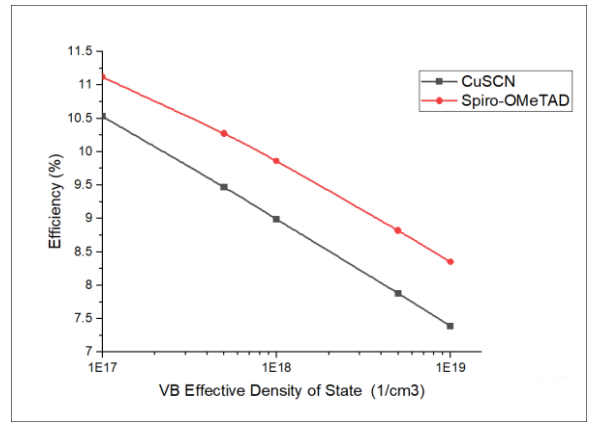

**Figure 7**: Efficiency vs VB effective density of state of the absorber layer for the best absorber layer thickness.

With the increasing $N_V$ of the absorber layer, the efficiency of the solar cell decreased for both cases. Best efficiency was found when the value of $N_V$ was taken as $10^{17}$ cm$^{-3}$ for both instances the same as the literature (Anwar, Mahbub et al. 2017). At $10^{17}$ cm$^{-3}$ of $N_V$ of the absorber layer, the solar cell had PCE of 10.53% and 11.12% for CuSCN (HTL) and spiro-OMeTAD (HTL) arrangement, respectively. With the increasing VB effective density of state, the number of holes increases at the absorber layer. At the same time, their possibility of taking part in $J_0$ also rises (Anwar, Mahbub et al. 2017). With the increasing $J_0$, $V_{OC}$ declines according to equation (1), and thus, PCE also falls.

### 3.4 Effect of Interfacial Trap Density of State

Trap density of the HTL/ MASnI$_3$ interfacial layer was varied between $10^{10}$ cm$^{-2}$ to $10^{16}$ cm$^{-2}$ and also MASnI$_3$/ ETL interfacial layer trap density was varied between $10^5$ cm$^{-2}$ to $10^{16}$ cm$^{-2}$ for the best absorber thickness to find the variation in PCE as shown in figure 8 for both configurations.

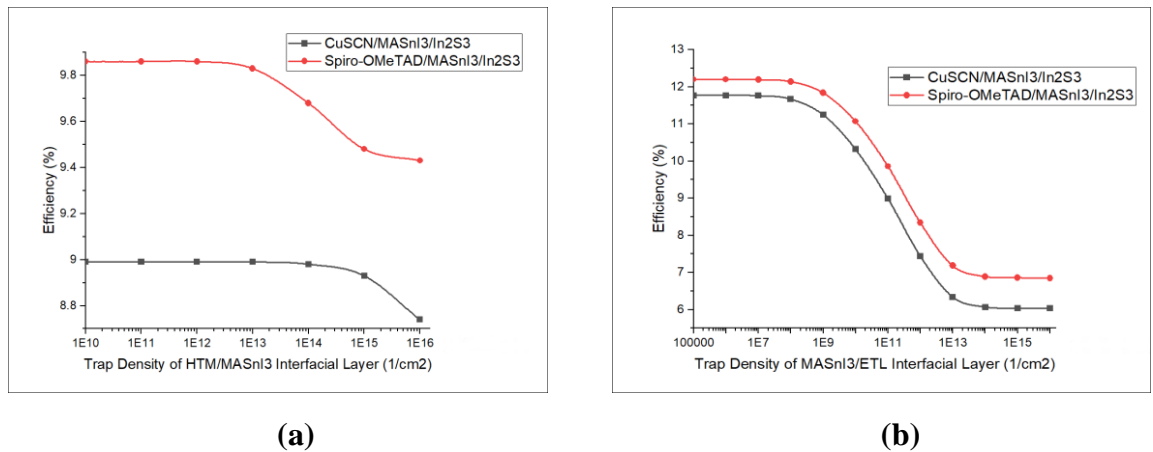

(a)                                (b)

**Figure 8**: Efficiency vs Trap Density for **(a)** HTL/ MASnI$_3$, **(b)** MASnI$_3$/ ETL Interfacial layers for the best absorber layer thickness.

With the increasing interfacial trap density, the efficiency of the solar cell generally followed a downward trend for both cases. For CuSCN/ MASnI$_3$ and spiro-OMeTAD/ MASnI$_3$ interfacial layers, there was no significant change in PCE when their trap densities were lower than $10^{13}$ cm$^{-2}$ (PCE 8.99%) and $10^{12}$ cm$^{-2}$ (PCE 9.86%) respectively which were taken as the optimum values, and these interfacial trap density values were close to the value found in the literature (Anwar, Mahbub et al. 2017). However, for MASnI$_3$/ ETL interfacial layers in both cases, the almost stable efficiency point was found at $10^8$ cm$^{-2}$ (PCE 11.67% for CuSCN and PCE 12.14% for spiro-OMeTAD arrangement) which was significantly lower from the value found for MASnI$_3$/ Zinc Oxide nanorod interfacial layer in the literature (Anwar, Mahbub et al. 2017). Increasing interfacial trap density decreases PCE because interface traps at the high level are the recombination centres, thus contributing to the enhancement in the shunt resistance (MaríSoucase, Pradas et al. 2016).

## *3.5 Effect of Bandgap of the Absorber Layer*

Tin based perovskite has a tunable bandgap between 1.3 eV to 2.15 eV (Mandadapu, Vedanayakam et al. 2017). In this simulation, the bandgap of the solar cell for both configurations was varied between 1.3 eV to 2.1 eV for the best absorber layer thickness to find the variation in efficiency, FF, $V_{OC}$ and $J_{SC}$ as shown in figure 9.

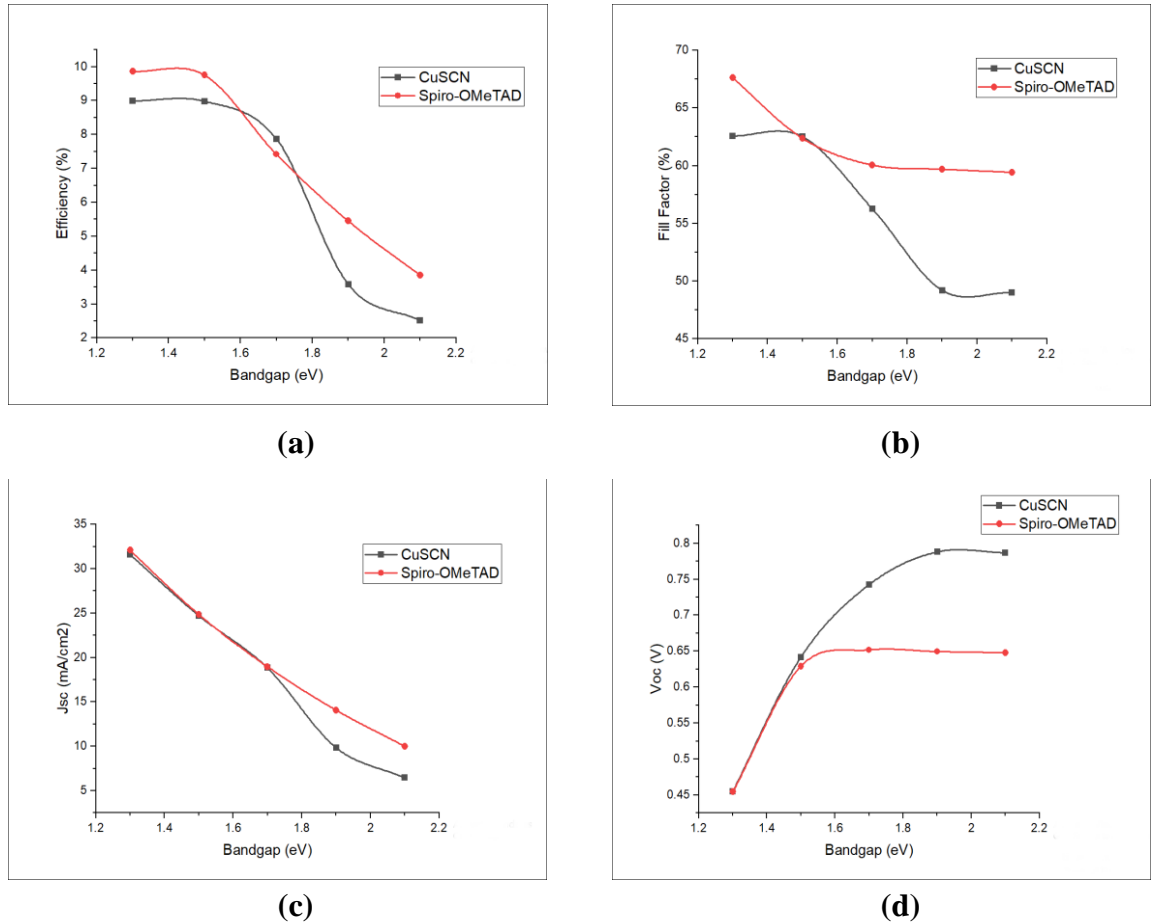

**Figure 9**: **(a)** PCE, **(b)** FF, **(c)** $J_{SC}$ and **(d)** $V_{OC}$ diagram with varying bandgap for the best absorber layer thickness.

With the increasing bandgap, PCE, FF, and $J_{SC}$ decreased, whereas $V_{OC}$ increased for both cases. $V_{OC}$ is a direct function of the bandgap, and higher bandgap leads to the higher $V_{OC}$ and lower rate of radiative recombination. $J_{SC}$ decreases with increasing bandgap due to less generation of electrons, as few photons have sufficient energy. Again, with increasing bandgap, FF reduces for the mismatch between the HTL and absorber layer. As PCE is the function of FF, $V_{OC}$ and $J_{SC}$, it decreased at the higher bandgap of the absorber layer (Mandadapu, Vedanayakam et al. 2017). So, the best performance was found at 1.3 eV bandgap for both configurations which was in good agreement with the literature (Iefanova, Adhikari et al. 2016, Mandadapu, Vedanayakam et al. 2017).

### 3.6 J-V characteristic Curve

J-V properties for solar cell with both configurations were investigated based upon the optimum values of input parameters and the output parameters are listed in table 2. At optimized condition, *SLG/ FTO/ In$_2$S$_3$/ MASnI$_3$/ Spiro-OMeTAD/ Au* solar cell exceeded the experimental efficiency of *SLG/ FTO/ In$_2$S$_3$ nanoflake/ MAPbI$_3$/ Spiro-OMeTAD/ Au* solar cell that had a PCE of 14.02% ($J_{SC}$ = 21.65 mA/cm$^2$, $V_{OC}$ = 1.02045 V and FF = 65%) (Yu, Zhao et al. 2019) and was near the experimental PCE of *SLG/ FTO/ In$_2$S$_3$/ MAPbI$_3$/ Spiro-OMeTAD/ Au* solar cell that had a PCE of 18.83% ($J_{SC}$ = 22.98 mA/cm$^2$, $V_{OC}$ = 1.10 V and FF = 75%) (Xu, Wu et al. 2018). Figure 10 shows J-V characteristics of optimized cell for both HTL arrangement. The working temperature was taken at 300K with standard illumination of AM1.5G solar spectrum.

**Table 2**: Comparison of output parameters for the solar cell with two alternate HTL configurations at the optimized condition.

| Serial | Device | $V_{OC}$ | $J_{SC}$ | Fill Factor | Efficiency |
|---|---|---|---|---|---|
| | | V | mA/cm2 | % | % |
| 1 | **CuSCN/ MASnI$_3$/ In$_2$S$_3$** | 0.7529 | 32.596613 | 75.19 | 18.45 |
| 2 | **Spiro-OMeTAD/ MASnI$_3$/ In$_2$S$_3$** | 0.7478 | 33.438154 | 77.28 | 19.32 |

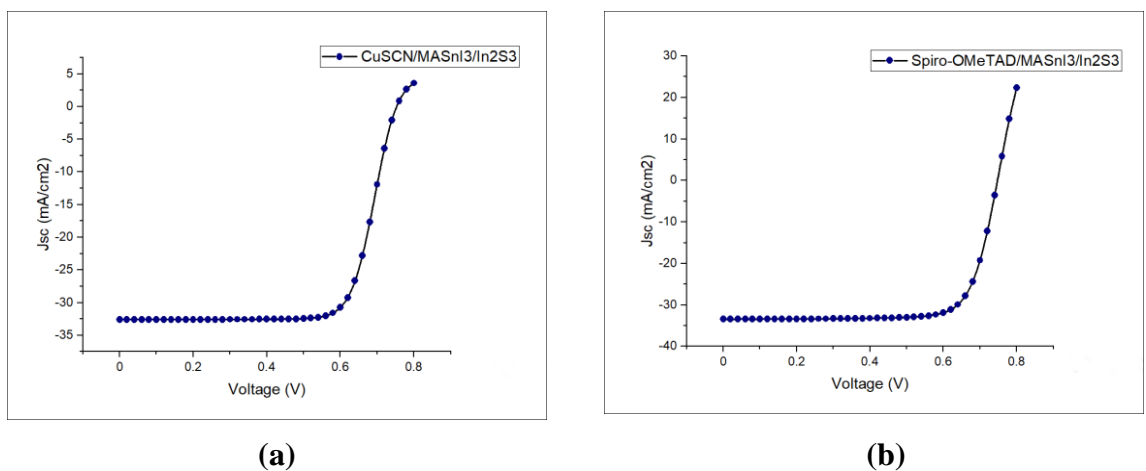

(a)      (b)

**Figure 10**: J-V characteristic curves of solar cell **(a)** CuSCN/ MASnI$_3$/ ETL and **(b)** Spiro-OMeTAD/ MASnI$_3$/ ETL arrangement at the optimized condition.

The EQE curves for the optimized condition for both configurations within the range between 400-1000 nm wavelength are given in figure 11, which follow similar trends as the literature (Niemegeers 2014). It was found that with the increasing wavelength from 400 nm, EQE decreased and at 960 nm wavelength of light, it became almost zero for both configurations.

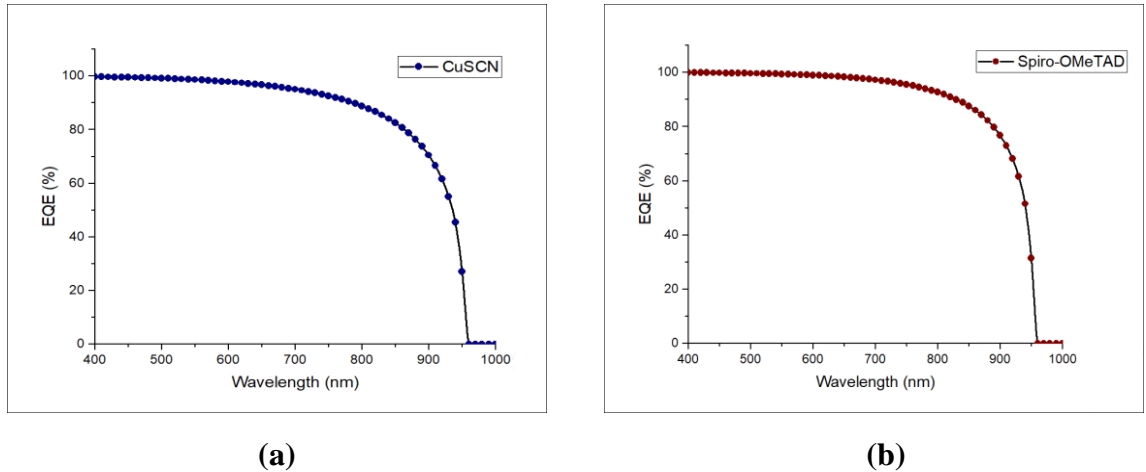

(a) (b)

**Figure 11**: EQE curves of solar cell **(a)** CuSCN/ MASnI$_3$/ ETL, **(b)** Spiro-OMeTAD/ MASnI$_3$/ ETL arrangement at optimized condition.

*3.7 Effect of Temperature*

Changing the temperature of the solar cell affects the overall cell performance. In the simulation, we kept the temperature fixed at 300K at first. Then it was varied from 300K to 500K to find the influence of working temperature on PCE, $V_{OC}$, $J_{SC}$ and FF for the best absorber thickness as showed in figure 12 for both configurations. It was found that with the increasing temperature, PCE, $V_{OC}$, $J_{SC}$ and FF of the solar cell decreased for both configurations, because the carrier concentrations, mobility of the charge carriers, resistance and bandgap of the materials alters at a higher temperature (MaríSoucase, Pradas et al. 2016).

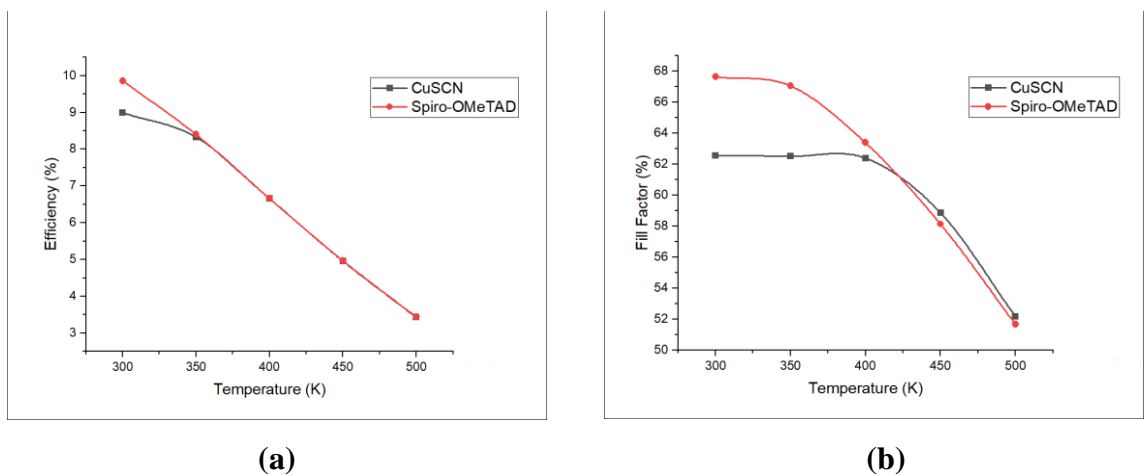

(a) (b)

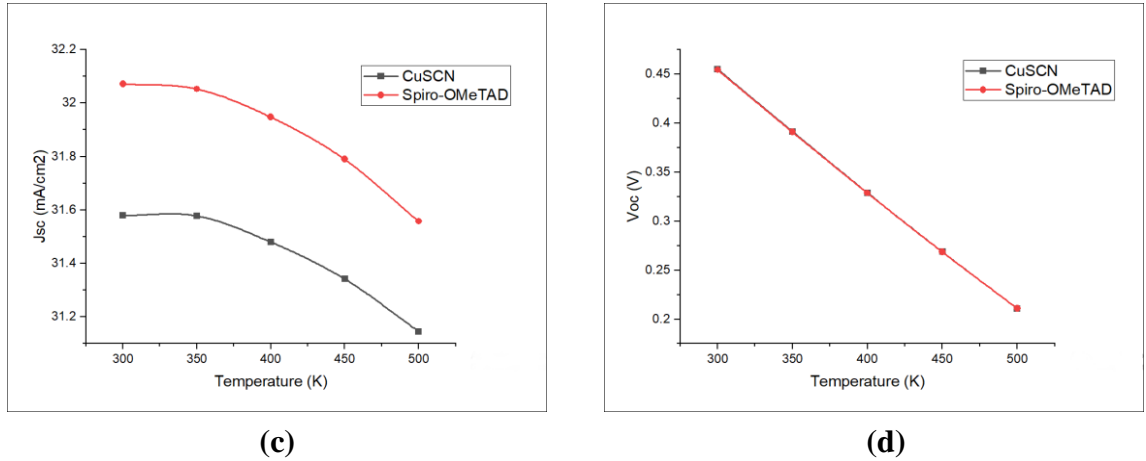

| (c) | (d) |

**Figure 12**: **(a)** PCE, **(b)** FF, **(c)** $J_{SC}$ and **(d)** $V_{OC}$ diagram with varying working temperature for best absorber layer thickness.

The efficiency dropped from 8.99% to 3.43% for CuSCN and from 9.86% to 3.44% for spiro-OMeTAD arrangement respectively at 500K temperature. Increased temperature causes a rise in stress and deformation, resulting in more interfacial defects and enhanced SRH recombination (Zandi, Saxena et al. 2020). Thus, diffusion length reduces and series resistance increases, which results in decreased PCE and FF (Mandadapu, Vedanayakam et al. 2017). The reduction in $J_{SC}$ could be due to enhanced bandgap with the increasing temperature and negative temperature coefficient for $J_{SC}$ which is applicable for the perovskite solar cells (Yu, Chen et al. 2011, Schwenzer, Rakocevic et al. 2018). Again, $J_0$ increases with the increasing temperature, and so, $V_{OC}$ gets lower according to equation (1) (Riedel, Parisi et al. 2004, Lin, Lin et al. 2014). Furthermore, electrons become unstable at a higher temperature in the solar cell due to the rise in energy causing enhanced recombination, resulting in low PCE (Mandadapu, Vedanayakam et al. 2017). So, 300K temperature was used as the working temperature for the optimal performance same as the literature (Lin, Lin et al. 2014, Mandadapu, Vedanayakam et al. 2017).

*3.8 Effect of Light Spectrum*

The simulations were done at the standard illumination of AM1.5G at first. Later, the wavelength was varied between 400 nm to 900 nm to find the effect of light spectrum on the efficiency as shown as figure 13 for both configurations. According to the literature, $MASnI_3$ perovskite can have light absorption onset up to 1050 nm (Lazemi, Asgharizadeh et al. 2018).

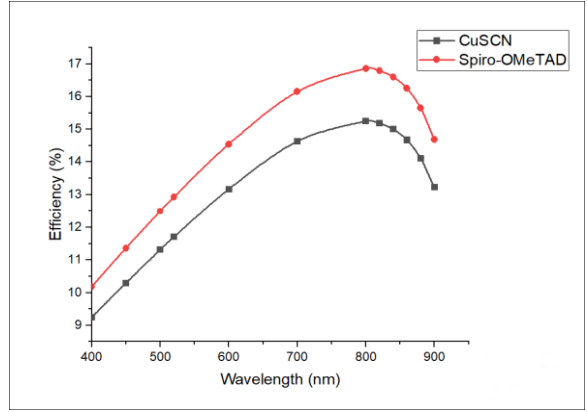

**Figure 13**: PCE diagram with the varying wavelength of light for the best absorber layer thickness.

Variation in PCE occurs with the change of wavelength, but the change is not always steady as materials are sensitive to different wavelength depending on their quantum characteristics (Ashraf and Alam 2020). It was found that with the increase of the wavelength, efficiencies of solar cell for both arrangements increased up to a certain point, then decreased. Output parameters showed that maximum PCE was found at 800 nm for both configurations. The efficiency increased up to 15.25% and 16.86% at 800 nm wavelength for CuSCN and spiro-OMeTAD configurations, respectively. The efficiency declined when the value of wavelength was lower than 800 nm due to enhanced recombination (Ashraf and Alam 2020).

*3.9 Effect of Band-to-Band Radiative Recombination Rate*

In the simulation, we increased band-to-band radiative recombination rate from the initial $3 \times 10^{-11}$ cm$^3$ sec$^{-1}$ to $3 \times 10^{-9}$ cm$^3$ sec$^{-1}$ as shown in figure 14 to observe the $V_{OC}$, $J_{SC}$, FF and PCE trends of the solar cell with both configurations and for each rate, perovskite layer thickness was varied from 200 nm to 800 nm. $V_{OC}$, $J_{SC}$, FF and PCE decreased with the increase of radiative recombination rate for both configurations.

For a solar cell, the total recombination rate can be expressed as a polynomial which is contingent on the charge carrier density (Wehrenfennig, Eperon et al. 2014), as shown by the following equation:

$$\frac{dn}{dt} = -k_3 n^3 - k_2 n^2 - k_1 n \qquad (2)$$

Where *n* (cm$^{-3}$) is the excess carrier density. Auger, band-to-band radiative and SRH recombination are expressed by the cubic, quadratic and linear terms, respectively.

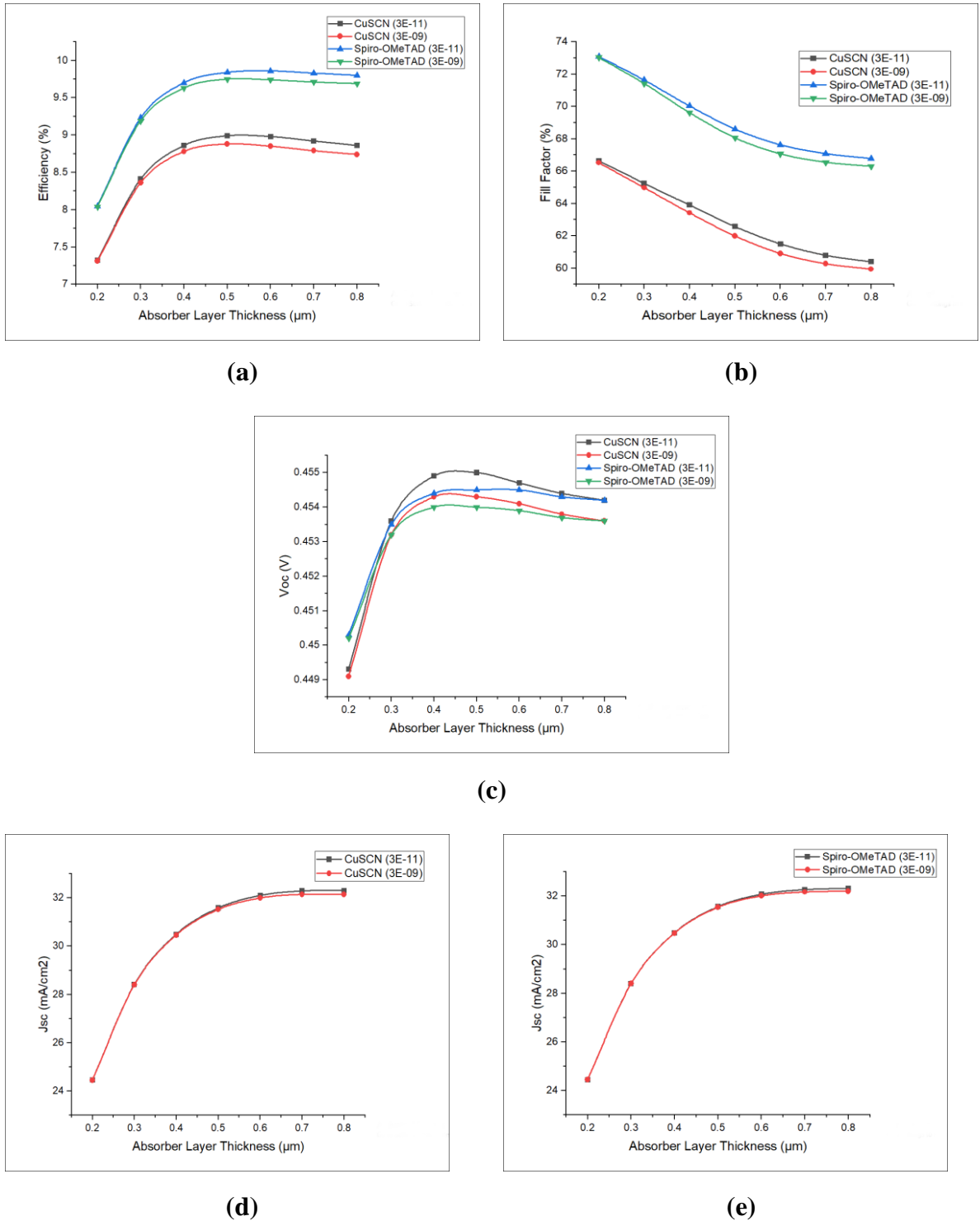

**Figure 14**: **(a)** PCE, **(b)** FF, **(c)** V$_{OC}$ for both configurations and **(d)** J$_{SC}$ (CuSCN configuration), **(e)** J$_{SC}$ (spiro-OMeTAD configuration) diagram with varying absorber layer thickness for different band-to-band recombination rates.

SCAPS deals with the band-to-band radiative and SRH recombination only, and they are also more significant. SRH recombination was discussed previously. Band-to-band radiative recombination, primarily a material property, is contingent on the instantaneous recombination of electrons and holes produced from electron-hole pairs generated because of light absorption. Because of this, it is the most intrinsic form of recombination, and furthermore, it depends on the perovskite layer's acceptor concentration ($N_a$) and donor concentrations ($N_d$) (Islam, Jani et al. 2020).

All the parameters in figure 14 followed the similar trends as in figure 2 for both configurations, which suggests that band-to-band radiative recombination is an intrinsic property depending on the absorber's $N_a$ and $N_d$. The effect of varying the recombination rate on $J_{SC}$ was small, which indicates that radiative recombination has minimal effect on the electron-hole pair generation and transport in the solar cell. Similarly, the negligible effect on $V_{OC}$ shows that radiative recombination rate has almost no effect on $V_{OC}$. However, the effect of radiative recombination on FF was slightly higher than the effects on $V_{OC}$ and $J_{SC}$. The decreasing FF with the increasing radiative recombination rate was because, electron-hole pairs transport through the cell less effectively with more losses via recombination (Bartesaghi, del Carmen Pérez et al. 2015, Islam, Jani et al. 2020). Again, at $3 \times 10^{-9}$ cm$^3$/sec radiative recombination rate and 500 nm of absorber thickness, the solar cell with CuSCN (HTL) had maximum PCE of 8.88%, and cell with spiro-OMeTAD (HTL) had maximum PCE of 9.75% at. So, when the band-to-band radiative recombination rate increased from the initial $3 \times 10^{-11}$ cm$^3$ sec$^{-1}$ to $3 \times 10^{-9}$ cm$^3$ sec$^{-1}$, there was about 0.11% decrease in the maximum PCE for both configurations.

## 4. Conclusion

In this work, SCAPS 1D was utilized to investigate the optimal behaviour of tin-based perovskite solar cell with two different HTL configurations. Under 300K working temperature and standard illumination of AM 1.5G, the absorber layer thickness was varied from 0.2 to 0.8 μm, and the optimum values for CuSCN (HTL) and spiro-OMeTAD (HTL) configuration were recorded as 0.5 μm and 0.6 μm, respectively. Again, at best absorber thickness, $N_t$ of absorber layers was varied between $10^{10}$ cm$^{-3}$ to $10^{18}$ cm$^{-3}$, the bandgap of the perovskite was changed from 1.3 eV to 2.1 eV, $N_v$ of the absorber layer was modified from $10^{17}$ cm$^{-3}$ to $10^{19}$ cm$^{-3}$ in both cell configurations. The device performed well when the $N_t$ of the absorber layer was kept at $10^{14}$ cm$^{-3}$ with $10^{17}$ cm$^{-3}$ of $N_v$ and perovskite bandgap of 1.3 eV in both cases. $N_t$ of ETL was varied between $10^{14}$ cm$^{-3}$ to $10^{18}$ cm$^{-3}$ and the optimum $N_t$ of ETL was recorded as $10^{15}$ cm$^{-3}$. Interfacial trap density of state was also varied from $10^{10}$ cm$^{-2}$ to $10^{16}$ cm$^{-2}$ for HTL/ MASnI$_3$ and $10^5$ cm$^{-2}$ to $10^{16}$ cm$^{-2}$ for MASnI$_3$/ ETL and the optimum trap densities for CuSCN/ MASnI$_3$, spiro-OMeTAD/ MASnI$_3$ and MASnI$_3$/ In$_2$S$_3$ interfacial layers were found as $10^{13}$ cm$^{-2}$, $10^{12}$ cm$^{-2}$ and $10^8$ cm$^{-2}$, respectively. When the optimal values of input parameters were considered, the highest efficiency achieved was achieved as 18.45% ($V_{oc}$ = 0.7529 V, $J_{sc}$ = 32.596613 mA/cm$^2$, FF = 75.19%) for CuSCN and 19.32% ($V_{oc}$ = 0.7478 V, $J_{sc}$ = 33.438154 mA/cm$^2$, FF = 77.28%) for spiro-OMeTAD configuration. So, when the cheaper CuSCN HTL was used, the PCE was decreased by 0.87% only and so, it can be a suitable alternative to the traditional and more expensive spiro-OMeTAD HTL. Again, In$_2$S$_3$ as ETL shows good PV performance that is comparable to the solar cells with TiO$_2$ ETL. The EQE curves were shown between 400-1000 nm wavelength for the best absorber thickness and defect density and also, for the optimized condition for both configurations. Moreover, the operating temperature was changed from 300K to 500K to see the changes in the PV parameters and the highest efficiency was found at 300K for both configurations. Again, the wavelength of light was varied from 400 nm to 900 nm to observe the influence of light spectrum on efficiency and highest efficiency was found at 800 nm wavelength for both cases. Furthermore, when the radiative recombination rate was increased from the initial $3 \times 10^{-11}$ cm$^3$ sec$^{-1}$ to $3 \times 10^{-9}$ cm$^3$ sec$^{-1}$, there was about 0.11% decrease in the maximum PCE for both configurations. All these simulation results will assist in replacing the commonly used toxic MAPbI$_3$ by the non-toxic MASnI$_3$ and ensure its high-efficiency performance after optimization that is almost equivalent to the lead-based cell when coupled with the In$_2$S$_3$ ETL. But in the real scenario, MASnI$_3$ will have lower stability than MAPbI$_3$, which will result in decreased PCE of the solar cell. So, we propose the addition of SnF$_2$ or SnCl$_2$ to improve the stability of our tin-based perovskite so that this high PCE can be attained. Finally, experimental studies are required for an extensive investigation to ensure the feasibility of our proposed structure.

## 5. Acknowledgement

The authors acknowledge Dr Marc Burgelman at the University of Gent, Belgium, for providing the simulation platform SCAPS 1-D.

## 6. Conflict of Interests

There was no conflict of interests.

# Appendix

MASnI$_3$ = Methylammonium Tin-Iodide (CH$_3$NH$_3$SnI$_3$)

MAPbI$_3$ = Methylammonium Lead-Iodide (CH$_3$NH$_3$SnI$_3$)

In$_2$S$_3$ = Indium (III) sulfide

TiO$_2$ = Titanium Oxide

CuSCN = Copper (I) thiocyanate

Spiro-OMeTAD = 2,2',7,7'-Tetrakis [N, N-di(4-methoxyphenyl) amino]-9,9'-spirobifluorene

Au = Gold

SLG = Soda-lime glass

FTO = Fluorine-doped Tin Oxide

ETL= Electron Transport Layer

HTL= Hole Transport Layer

FF = Field Factor (%)

PCE = Photoconversion Efficiency (%)

V$_{OC}$ = Open-Circuit Voltage (V)

J$_{SC}$ = Short-Circuit Current Density (mA/cm$^2$)

J$_0$ = Reverse Saturation Current Density (mA/cm$^2$)

EQE = External Quantum Efficiency (%)

Eg = Bandgap (eV)

α(hv) = Optical Absorption Constant (cm$^{-1}$)

SRH = Shockley-Read-Hall

N$_t$ = Defect Density (cm$^{-3}$)

N$_v$ = VB Effective Density of States (cm$^{-3}$)

N$_a$ = Acceptor Concentration (cm$^{-3}$)

N$_d$ = Donor Concentration (cm$^{-3}$)

n = Excess Carrier Density (cm$^{-3}$)

*A* = Ideality Factor

q = Elementary Charge

$\frac{K_B T}{q}$ = Thermal Voltage (V)

# References

Anwar, F., R. Mahbub, S. S. Satter and S. M. Ullah (2017). "Effect of different HTM layers and electrical parameters on ZnO nanorod-based lead-free perovskite solar cell for high-efficiency performance." International Journal of Photoenergy **2017**: 9. doi: 10.1155/2017/9846310

Aouaj, M. A., R. Diaz, A. Belayachi, F. Rueda and M. Abd-Lefdil (2009). "Comparative study of ITO and FTO thin films grown by spray pyrolysis." Materials Research Bulletin **44**(7): 1458-1461.

Arora, N., M. I. Dar, A. Hinderhofer, N. Pellet, F. Schreiber, S. M. Zakeeruddin and M. Grätzel (2017). "Perovskite solar cells with CuSCN hole extraction layers yield stabilized efficiencies greater than 20%." Science **358**(6364): 768-771.

Ashraf and Alam (2020). "Numerical simulation of CIGS, CISSe and CZTS-based solar cells with In2S3 as buffer layer and Au as back contact using SCAPS 1-D." Engineering Research Express: 21-22. doi: 10.1088/2631-8695/abade6

Bach, U., D. Lupo, P. Comte, J.-E. Moser, F. Weissörtel, J. Salbeck, H. Spreitzer and M. Grätzel (1998). "Solid-state dye-sensitized mesoporous TiO2 solar cells with high photon-to-electron conversion efficiencies." Nature **395**(6702): 583-585.

Baig, F., Y. H. Khattak, B. Marí, S. Beg, A. Ahmed and K. Khan (2018). "Efficiency Enhancement of CH3NH3 SnI3 Solar Cells by Device Modeling." Journal of Electronic Materials **47**(9): 5275-5282.

Bangari, N., V. K. Singh and V. K. Sharma (2020). "Experimental investigation of thin-film solar cells as a wearable power source." Energy Sources, Part A: Recovery, Utilization, and Environmental Effects: 1-21. doi: 10.1080/15567036.2020.1776794

Bartesaghi, D., I. del Carmen Pérez, J. Kniepert, S. Roland, M. Turbiez, D. Neher and L. J. A. Koster (2015). "Competition between recombination and extraction of free charges determines the fill factor of organic solar cells." Nature communications **6**(1): 1-10.

Burgelman, M., P. Nollet and S. Degrave (2000). "Modelling polycrystalline semiconductor solar cells." Thin solid films **361**: 527-532.

Chen, Z., B. Turedi, A. Y. Alsalloum, C. Yang, X. Zheng, I. Gereige, A. AlSaggaf, O. F. Mohammed and O. M. Bakr (2019). "Single-crystal MAPbI3 perovskite solar cells exceeding 21% power conversion efficiency." ACS Energy Letters **4**(6): 1258-1259.

De Los Santos, I. M., H. J. Cortina-Marrero, M. Ruíz-Sánchez, L. Hechavarría-Difur, F. Sánchez-Rodríguez, M. Courel and H. Hu (2020). "Optimization of CH3NH3PbI3 perovskite solar cells: A theoretical and experimental study." Solar Energy **199**: 198-205.

Du, H.-J., W.-C. Wang and J.-Z. Zhu (2016). "Device simulation of lead-free CH3NH3SnI3 perovskite solar cells with high efficiency." Chinese Physics B **25**(10): 108802.

Fantacci, S., F. De Angelis, M. K. Nazeeruddin and M. Grätzel (2011). "Electronic and optical properties of the spiro-MeOTAD hole conductor in its neutral and oxidized forms: a DFT/TDDFT investigation." The Journal of Physical Chemistry C **115**(46): 23126-23133.

Hossain, M. I. (2012). "Fabrication and characterization of CIGS solar cells with In2S3 buffer layer deposited by PVD technique." Chalcogenide Letters **9**(5): 185-191.

Hossain, M. I., P. Chelvanathan, M. Zaman, M. Karim, M. Alghoul and N. Amin (2011). "Prospects of indium sulphide as an alternative to cadmium sulphide buffer layer in CIS based solar cells from numerical analysis." Chalcogenide Lett **8**(5): 315-324.

Hou, Y., X. Chen, S. Yang, Y. L. Zhong, C. Li, H. Zhao and H. G. Yang (2017). "Low-temperature processed In2S3 electron transport layer for efficient hybrid perovskite solar cells." Nano Energy **36**: 102-109.

Huang, L.-j., N.-f. Ren, B.-j. Li and M. Zhou (2014). "Improvement in overall photoelectric properties of Ag/FTO bilayer thin films using furnace/laser dual annealing." Materials Letters **116**: 405-407.

Husainat, A., W. Ali, P. Cofie, J. Attia, J. Fuller and A. Darwish (2020). "Simulation and Analysis Method of Different Back Metals Contact of CH3NH3PbI3 Perovskite Solar Cell Along with Electron Transport Layer TiO2 Using MBMT-MAPLE/PLD." American Journal of Optics and Photonics **8**(1): 6-26.

Iefanova, A., N. Adhikari, A. Dubey, D. Khatiwada and Q. Qiao (2016). "Lead free CH3NH3SnI3 perovskite thin-film with p-type semiconducting nature and metal-like conductivity." AIP Advances **6**(8): 085312.

Islam, M. T., M. R. Jani, S. M. Al Amin, M. S. U. Sami, K. M. Shorowordi, M. I. Hossain, M. Devgun, S. Chowdhury, S. Banerje and S. Ahmed (2020). "Numerical simulation studies of a fully inorganic Cs2AgBiBr6 perovskite solar device." Optical Materials **105**: 109957.

Jeyakumar, R., A. Bag, R. Nekovei and R. Radhakrishnan (2020). "Influence of Electron Transport Layer (TiO2) Thickness and Its Doping Density on the Performance of CH3NH3PbI3-Based Planar Perovskite Solar Cells." Journal of Electronic Materials **49**(6): 1-7.

Kharaji Manouchehrabadi, M. and S. Yaghoubi (2020). "A game theoretic incentive model for closed-loop solar cell supply chain by considering government role." Energy Sources, Part A: Recovery, Utilization, and Environmental Effects: 1-25. doi: 10.1080/15567036.2020.1764150

Khelifi, S., J. Verschraegen, M. Burgelman and A. Belghachi (2008). "Numerical simulation of the impurity photovoltaic effect in silicon solar cells." Renewable Energy **33**(2): 293-298.

Lan, Z., J. Wu, J. Lin and M. Huang (2014). "A Highly Efficient Dye-sensitized Solar Cell with a Blocking Layer and TiCl4 Treatment to Suppress Dark Reaction." Energy Sources, Part A: Recovery, Utilization, and Environmental Effects **36**(16): 1810-1817. doi: 10.1080/15567036.2011.561276

Lazemi, M., S. Asgharizadeh and S. Bellucci (2018). "A computational approach to interface engineering of lead-free CH3NH3SnI3 highly-efficient perovskite solar cells." Physical Chemistry Chemical Physics **20**(40): 25683-25692.

Lee, B., J. He, R. P. Chang and M. G. Kanatzidis (2012). "All-solid-state dye-sensitized solar cells with high efficiency." Nature **485**(7399): 486-489.

Lenka, T., A. Soibam, K. Dey, T. Maung and F. Lin (2020). "Numerical analysis of high-efficiency lead-free perovskite solar cell with NiO as hole transport material and PCBM as electron transport material." CSI Transactions on ICT **8**: 111-116. doi: 10.1007/s40012-020-00291-7

Li, F., C. Zhang, J.-H. Huang, H. Fan, H. Wang, P. Wang, C. Zhan, C.-M. Liu, X. Li, L.-M. Yang, Y. Song and K.-J. Jiang (2019). "A Cation-Exchange Approach for the Fabrication of Efficient Methylammonium Tin Iodide Perovskite Solar Cells." Angewandte Chemie International Edition **58**(20): 6688-6692.

Lin, P., L. Lin, J. Yu, S. Cheng, P. Lu and Q. Zheng (2014). "Numerical simulation of Cu2ZnSnS4 based solar cells with In2S3 buffer layers by SCAPS-1D." Journal of Applied Science and Engineering **17**(4): 383-390.

Mandadapu, U., S. V. Vedanayakam, K. Thyagarajan, M. R. Reddy and B. J. Babu (2017). "Design and simulation of high efficiency tin halide perovskite solar cell." International Journal of Renewable Energy Research **7**(4): 1603-1612.

MaríSoucase, B., I. G. Pradas and K. R. Adhikari (2016). "Numerical Simulations on Perovskite Photovoltaic Devices." Perovskite Materials: Synthesis, Characterisation, Properties, and Applications InTechOpen, London, Ch.15-16.

Marlein, J., K. Decock and M. Burgelman (2009). "Analysis of electrical properties of CIGSSe and Cd-free buffer CIGSSe solar cells." Thin solid films **517**(7): 2353-2356.

Meng, Q., Y. Chen, Y. Y. Xiao, J. Sun, X. Zhang, C. B. Han, H. Gao, Y. Zhang and H. Yan (2020). "Effect of temperature on the performance of perovskite solar cells." Journal of Materials Science: Materials in Electronics: 1-9. doi: 10.1007/s10854-020-03029-y.

Niemegeers, A. (2014). "Models for the optical absorption α(λ) of materials in SCAPS." (University of Gent).

Niemegeers, A., M. Burgelman, K. Decock, J. Verschraegen and S. Degrave (2014). "SCAPS manual." University of Gent.

Nine, K. B., M. F. Hossain and S. A. Mahmood (2019). Analysis of Stable, Environment Friendly and Highly Efficient Perovskite Solar Cell. TENCON 2019-2019 IEEE Region 10 Conference (TENCON), Kochi, India, 2019: 1825-1828. doi: 10.1109/TENCON.2019.8929358.

Rai, S., B. Pandey and D. Dwivedi (2020). Device simulation of low cost HTM free perovskite solar cell based on TiO2 electron transport layer. AIP Conference Proceedings, AIP Publishing LLC **2220**(1): 140022.

Riedel, I., J. Parisi, V. Dyakonov, L. Lutsen, D. Vanderzande and J. C. Hummelen (2004). "Effect of temperature and illumination on the electrical characteristics of polymer–fullerene bulk-heterojunction solar cells." Advanced Functional Materials **14**(1): 38-44.

Schwenzer, J. A., L. Rakocevic, R. Gehlhaar, T. Abzieher, S. Gharibzadeh, S. Moghadamzadeh, A. Quintilla, B. S. Richards, U. Lemmer and U. W. Paetzold (2018). "Temperature variation-induced performance decline of perovskite solar cells." ACS applied materials & interfaces **10**(19): 16390-16399.

Sharma, R. M. (2019). "Perovskite Solar Cell design using Tin Halide and Cuprous Thiocyanate for Enhanced Efficiency." International Journal of Engineering and Advanced Technology **8**(16): 2817-2825. doi: 10.35940/ijeat.F8778.088619

Tai, Q., J. Cao, T. Wang and F. Yan (2019). "Recent advances toward efficient and stable tin-based perovskite solar cells." EcoMat **1**(1): e12004.

Wehrenfennig, C., G. E. Eperon, M. B. Johnston, H. J. Snaith and L. M. Herz (2014). "High charge carrier mobilities and lifetimes in organolead trihalide perovskites." Advanced materials **26**(10): 1584-1589.

Wetzelaer, G. J. A., M. Scheepers, A. M. Sempere, C. Momblona, J. Ávila and H. J. Bolink (2015). "Trap-assisted non-radiative recombination in organic–inorganic perovskite solar cells." Advanced Materials **27**(11): 1837-1841.

Xu, Z., J. Wu, Y. Yang, Z. Lan and J. Lin (2018). "High-efficiency planar hybrid perovskite solar cells using indium sulfide as electron transport layer." ACS Applied Energy Materials **1**(8): 4050-4056.

Yang, B., M. Wang, X. Hu, T. Zhou and Z. Zang (2019). "Highly efficient semitransparent CsPbIBr2 perovskite solar cells via low-temperature processed In2S3 as electron-transport-layer." Nano Energy **57**: 718-727.

Yu, C., Z. Chen, J. J. Wang, W. Pfenninger, N. Vockic, J. T. Kenney and K. Shum (2011). "Temperature dependence of the band gap of perovskite semiconductor compound CsSnI3." Journal of Applied Physics **110**(6): 063526.

Yu, F., W. Zhao and S. F. Liu (2019). "A straightforward chemical approach for excellent In2S3 electron transport layer for high-efficiency perovskite solar cells." RSC advances **9**(2): 884-890.

Zandi, S., P. Saxena and N. E. Gorji (2020). "Numerical simulation of heat distribution in RGO-contacted perovskite solar cells using COMSOL." Solar Energy **197**: 105-110.

Zhou, D., T. Zhou, Y. Tian, X. Zhu and Y. Tu (2018). "Perovskite-based solar cells: materials, methods, and future perspectives." Journal of Nanomaterials **2018**: 15. doi: 10.1155/2018/8148072